\documentclass[default,iicol]{sn-jnl}

\usepackage{graphicx}
\usepackage{multirow}
\usepackage{amsmath,amssymb,amsfonts}
\usepackage{amsthm}
\usepackage{mathrsfs}
\usepackage[title]{appendix}
\usepackage{xcolor}
\usepackage{textcomp}
\usepackage{manyfoot}
\usepackage{booktabs}
\usepackage{algorithm}
\usepackage{algorithmicx}
\usepackage{algpseudocode}
\usepackage{listings}
\usepackage{subcaption}
\usepackage{circuitikz}
\usepackage{stfloats}   
\usepackage{balance}
\usepackage{cuted}
\usepackage{flushend}
\usepackage[numbers]{natbib}

\begin{document}

\title[Article Title]{Field Free Novel Architecture for Spintronic Flash Analog to Digital Converter}


\author*[1]{\fnm{Abin} \sur{Francis}}\email{abinfranciscl@gmail.com}

\author[1]{\fnm{Nikhil} \sur{Kumar}}\email{nikhilkumarcs@nitc.ac.in}

\author[2]{\fnm{Prince} \sur{Philip}}\email{princephilip@iisc.ac.in}

\affil*[1]{\orgdiv{Department of Electronics and Communication Engineering}, \orgname{National Institute of Technology Calicut}, \orgaddress{\street{Kattangal}, \city{Calicut}, \postcode{673601}, \state{Kerala}, \country{India}}}

\affil[2]{\orgdiv{Department of Electronic Systems Engineering}, \orgname{Indian Institute of Science}, \orgaddress{\street{CV Raman Road}, \city{Bengalore}, \postcode{10587}, \state{Karnataka}, \country{India}}}
%


\abstract{A 3-bit Analog-to-Digital Converter (ADC) is designed using perpendicular Spin-Orbit Torque Magnetic Tunnel Junction (SOT-MTJ). A sampled analog input signal is transmitted as a spin-orbit torque current (I$_{in}$) to a perpendicular SOT-MTJ, and deterministic switching is supported by the Voltage Controlled Magnetic Anisotropy (VCMA) and Spin Transfer Torque (STT) switching methods. Analog to digital conversion is done by comparing input signal with varied critical current of SOT-MTJ's. The critical current of each is SOT-MTJ governed by varying widths of Heavy Metal (HM). In the 3-bit ADC, there are two sets of 7 SOT-MTJs for quantizing input value, a conversion set and dummy set for comparing the change in resistance state.  As input signal passed through conversion set SOT-MTJ's switches from Parallel (P) to Anti-Parallel (AP) state if the input signal exceeds its critical current. The conversion set change in state is converted to thermometer codes by StrongARM latch comparator by comparing the resistance with dummy set SOT-MTJ's, where all the in P state or low resistance. A novel architecture is proposed for increasing speed of throughput, by utilizing the dummy set of as a conversion set and conversion set as dummy set, thus eliminating the reset step from analog to digital conversion. And by improving SOT-MTJ and timing blocks a field-free spin flash ADC has a power consumption of 476 $\mu$W with a conversion rate of 304.1 MS/s is produced.}

\maketitle



\vspace*{-1cm}
\section{Introduction}
In high-speed data acquisition, Flash ADCs stand as the workhorses of the industry, enabling rapid and precise conversion of continuous analog signals into digital representations. Demand in wearable devices, in-memory signal processing\cite{inmemory}, and neuromorphic computing\cite{neuromorphic} push for lower-footprint devices. The scaling down deep to submicron CMOS technology brings a handle full of effects, mainly static power consumption, threshold variations, and other short channel effects and process variations, implying CMOS is at saturation and deteriorates while going beyond\cite{ADC1}. The nonvolatile memory devices such as spintronics devices and resistive random access memory\cite{rram} (RRAM) with zero standby leakage current and endurance set a new path beyond CMOS.

Spinroncis device Magnetic Tunnel Junction (MTJ) is a good candidate for beyond CMOS. MTJ offers tunnel magneto-resistance (TMR) and tunnel magneto-capacitance (TMC) properties. As the magnetic moment, orientation in the free layer (FL) of MTJ can have two states, antiparallel(AP) and parallel (P) state, by Reference layer (RL) magnetic moment orientation. TMR and TMC vary according to states in MTJ. Hybrid circuits using CMOS and spintronics devices report the possibility of implementing Flash ADC with lower area and power consumption than typical CMOS Flash ADC \cite{spinADC1,spinADC2,spinADC3,spinADC4,spinADC5,spinADC6,spinADC7,spinADC8}. In \cite{spinADC1}, a 3-bit Flash ADC is proposed by using a resistor ladder and eight MTJ comparators controlled by Voltage Controlled Magnetic Anisotropy (VCMA) and Spin Hall Effect (SHE) switching mechanisms through a 2$\times$1 MUX without consideration of thermal noise. By eliminating the resistor ladder in Flash ADC, using variable width of Heavy Meatl(HM) in Spin-Orbit Transfer (SOT) MTJ controlled by SHE and Spin Transfer Torque (STT) mechanisums\cite{spinADC2} proposed spin-based Flash ADC. With improved read reliability using a binary search algorithm\cite{spinADC3}, they proposed 3-bit and 4-bit spin-based Flash ADC. 

In \cite{spinADC5}, a dual spin-based ADC with a 4-terminal MTJ was proposed to ensure more reliable quantization by double-checking estimation, producing a conversion rate of 667 MS/s. In \cite{spinADC7}, proposed a Magnetoelectric Spin-Orbit (MESO) device, by varying thickness of MESO and similarly varying width of HM, produced a conversion rate of 8.3GS/s and consumption of 0.14 pJ energy, with settling errors. In [7], the proposed spin ADC with variable width HM and read-out circuit using StrongARM latch comparator produced a conversion rate of 102MS/s with 416$\mu$W. In \cite{spinADC8}, experimentally validated quantization of analog signal to digital using in-plane MTJ(iMTJ). 

  The spin-based ADC designs discussed above lack a capacitance model and need to find a replacement for an external magnetic field for a more reliable design. In toggle MRAM by Everspin technology\cite{ToogleMRAM}, the external field is produced by passing current through the interconnect, thus producing a stray field. The MTJ read reliability issues are discussed in \cite{SpinADC9NAND}, where a magnetic field is produced as current flows through the interconnects. Controlling stray fields is more complex\cite{strayfield} and consumes power\cite{ToogleMRAM} when compared to the use of the Synthetic Antiferromagnetic(SAF) layer\cite{SAF} or Magnetic Hard Mask (MHM) layer\cite{MHM}, a field free technique. The conversion steps in the above discussed Spin Flash ADC have three steps, which can be reduced by modifying the circuits with lower power methods.  \begin{figure}[!t]
	\includegraphics[width=\columnwidth]{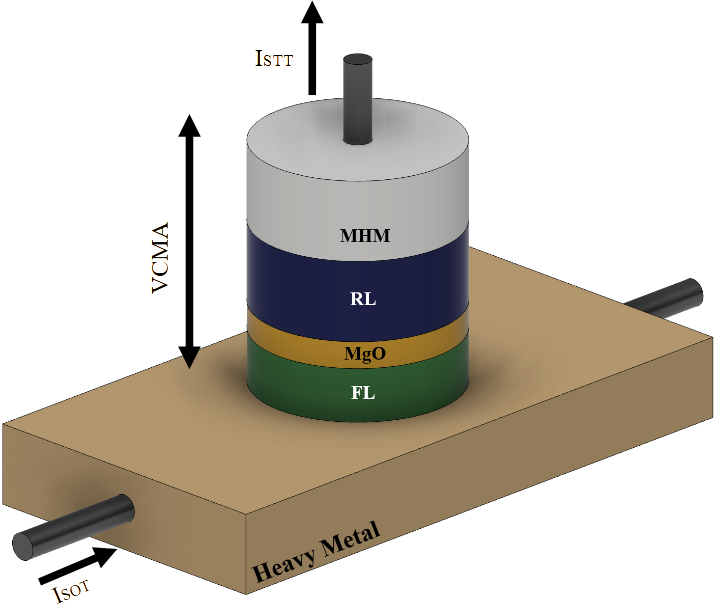}
	\caption{Improved SOT-MTJ which acts a quantizer using MHM.}\label{sot_improved_image}
\end{figure}

In this article, field-free deterministic switching of SOT perpendicular Magnetic Tunnel Junction(pMTJ), a three-terminal MTJ, is taken as a quantizer for conversion of analog to digital conversion. Fig. \ref{sot_improved_image}, represent SOT-MTJ which is composed of a Heavy Metal(HM) sandwiched in between two terminal MTJ, which have a 175\% TMR ratio\cite{SpinADC9NAND} and a layer of SAF layer\cite{SAF} or MHM layer\cite{MHM} is used for producing an in-plane magnetic field. 
\par A 3-bit ADC is designed  using 7 pMTJ acts as a quantizer, each pMTJ has its critical current by varying widths of HM. When the input current is passed through HM, each pMTJ compares with its critical current by SHE. The pMTJ switching current is supported by STT and voltage biasing to complete deterministic switching. The pMTJ switches from AP to P state, thus varying resistance, read out by the latched comparator and producing a thermometer code. The thermal noise effects in MTJ due to voltage bias are incorporated in the pMTJ model, making the device more practical. The proposed architecture for spin flash ADC produces a faster conversion rate and reduced power consumption. Utilizing the dummy set of 3-bit SOT-MTJ for conversion reduces the number of conversion steps in conventional spin Flash ADC. Section \ref{section2} shows the device modelling for the quantization of the input signal and the defining critical current for the quantizer. Section \ref{section3} shows the proposed architecture of the Spin Flash ADC. Section \ref{section4} shows the simulation results and comparison of proposed and conventional architecture. 
\section{Spin Quantizer}\label{section2}
Spin quantizing is the part where the first step of conversion occurs, where the SOT-MTJ switches from low resistance (P state) to high resistance (AP state). The resistance change is sensed by charge steering StrongARM latch comparator\cite{strongARM}, where it amplifies the change and converts it into thermometer code, using a priority encoder the thermometer code can be converted into binary code. SOT-MTJ model is developed under macromagnetic simulations, where the free layer's dynamics are considered as a whole single magnetic moment. The parameters used for simulations are discussed in table 
\par
\begin{figure}[!t]
	\includegraphics[height=0.8\columnwidth,width=\columnwidth]{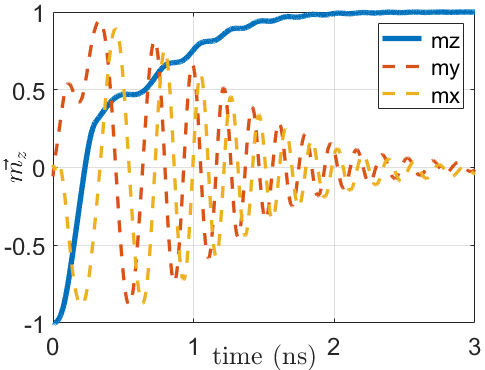}
	\caption{Single domain approximation simulation result of magnetic moment ($\vec{\mathbf{m}}$) in the free layer is shown, where its initial position is at (0,0,-1) in XYZ axis. The dynamics are changed by applying spin current through T2 to T4, at 0.7ns, aligning XY-plane parallel to spin current. Then with VCMA, a deterministic switching is made at 3ns and the final position of $\vec{\mathbf{m}}$ is at (0,0,1).}\label{fig2}
\end{figure}

 The p-MTJ is used with a heavy metal attached to FL and MHM is attached to the above RL as shown in Fig. \ref{sot_improved_image}. The purpose of MHM is to produce an in-plane magnetic field which helps in SHE. The device dynamics are calculated using the Landau-Lifshitz-Gilbert (LLG) equation which is: 
\begin{equation}\label{llgs}
	\frac{d\mathbf{m}}{dt} = -\gamma \mathbf{m} \times \mathbf{H}_{\text{eff}} + \alpha \mathbf{m} \times \frac{d\mathbf{m}}{dt} - \gamma \boldsymbol{\tau} 
\end{equation}
$\boldsymbol{\tau}$ consists of Spin Transfer Torque (STT) and Spin-Orbit Torque (SOT) terms, which control the FL dynamics.	The input signal is SOT terms and STT terms are for assisting the SOT in making switching deterministic.
\begin{equation}
	\label{LLGSTT_taw}
	\boldsymbol{\tau} = \boldsymbol{\tau}_{\mathrm{STT}} +\boldsymbol{\tau}_{\mathrm{SOT}}
\end{equation}The effective magnetic field ($\mathbf{H}_{\text{eff}}$) consists of Perpendicular Magnetic Anisotropy Field ($\mathbf{H}_{\text{pma}}$), Voltage-Controlled Magnetic Anisotropy ($\mathbf{H}_{\text{vcma}}$), Thermal Noise Field($\mathbf{H}_{\text{th}}$), In-plane magnetic field($\mathbf{H}_{\text{In-plane}}$), Demagnetization Field ($\mathbf{H}_{\text{d}}$). The SOT-MTJ model\cite{Voltagegatedmodel} is simulated in MATLAB and Cadence Virtuoso, where deterministic switching is observed under 2ns as shown in Fig. \ref{fig2}.
\begin{equation}
	\mathbf{H}_{\text{eff}} = \mathbf{H}_{\text{pma}} + \mathbf{H}_{\text{vcma}} + \mathbf{H}_{\text{d}} + \mathbf{H}_{\text{In-plane}} + \mathbf{H}_{\text{th}}
\end{equation}
The minimum current to switch the dynamics of FL is termed as Critcal Current ($I_c$), which as as the reference current when compared to current mode Flash ADC. Analytically after solving for $I_c$ in (1), we get as \cite{eqIc}
\begin{equation}
	I_{\mathrm {C}}= \frac {q\mu _{0}t_{\mathrm {FL}}M_{\mathrm {s}}{K_i}t_{\mathrm {HM}}\omega_{\mathrm {HM}}}{\hslash \xi \eta }
\end{equation} 	
$I_c$ plays a crucial role in quantizing input signal, which is adjusted indirectly by Voltage biasing. Voltage biasing also have the potential to change the effective magnetic field. MTJs Critical current can be varying $\omega_{\mathrm {HM}}$, which thickness of heavy metal. The biasing SOT-MTJ produces VCMA effect, which lowers the energy barrier. 
\begin{equation}
	\Delta \left({{V_b}} \right) = {E_b}\left({{V_b}} \right)/{k_B}T = \Delta \left(0 \right) - \ \xi A{V_b}/{k_B}T_0{t_{\rm ox}}\label{eq5}
\end{equation}
\par In the recent research \cite{thermaleffectsmodel,tempeffects} shows that under voltage biasing parameters such as Magnetic Saturation (Ms) and Interfacial Magnetic Anisotropy (Ki) is also affected which are repersented in eq. (\ref{eq6}) and (\ref{eq7}) $\eta, \xi$ are the fitting factor which are taken to follow the experiments result graph \cite{thermaleffectsmodel,tempeffects}.  The concurrent action of a SOT pulse and an MTJ bias pulse allows for reducing the critical switching energy below the level typical of spin-transfer-torque while preserving the ability to switch the MTJ on the sub-nanosecond time scale. $\Delta \left({{V_b}} \right)$ shows in the band gap under voltage bias shown in eq.(\ref{eq5}).
\begin{figure}[!t]
	\includegraphics[width=1\columnwidth]{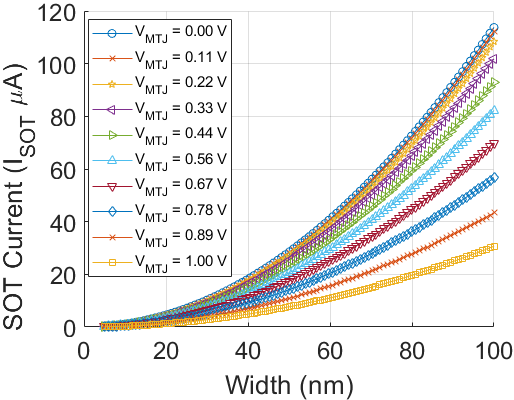}
	\caption{Simulation waveforms of SOT-MTJ critical current, varying with respect to width and voltage biasing.}\label{fig3}
\end{figure}
\begin{equation}
\small
	M_S\left(V_b\right)=M_S(0)\left\{1-\left[\left(T_0+k\left|V_b\right|^2\right) / T_C\right]^{\xi}\right\}\label{eq6}
\end{equation}

\begin{equation}
	\small 
    K_i\left(V_b\right)=K_i(0)\left\{1-\left[\left(T_0+k\left|V_b\right|^2\right) / T_C\right]^{\xi}\right\}^\eta\label{eq7}
\end{equation}
\par By performing simulation in MATLAB and Cadence Virtuoso three effects arising from the MTJ bias: the voltage-controlled change of the perpendicular magnetic anisotropy, current-induced heating, and the spin-transfer torque have been executed with the help of experiments\cite{interplay}. The experiments show the impact of voltage biasing on the temperature of the device and are also incorporated in the device model and variation of critical current of MTJ is shown Fig. \ref{fig3}.   
 \par
\begin{figure}[!t]
	\includegraphics[width=1\columnwidth]{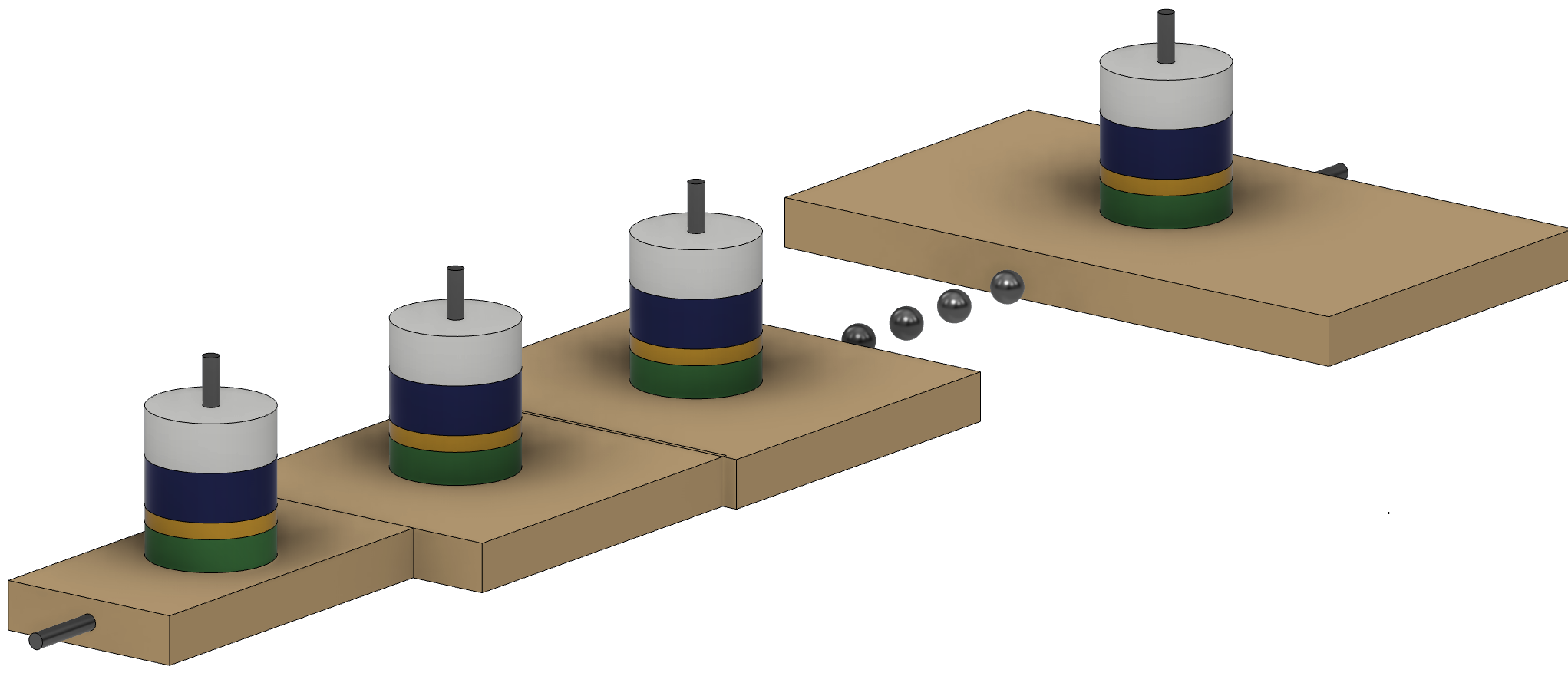}
	\caption{3-bit improved SOT-MTJ, which acts a 7 quantizer with increasing width for varying the critical current of SOT-MTJ .}\label{fif4}
\end{figure}

 The voltage biasing of MTJ ($V_{MTJ}$) helps and degrade the switching of MTJ. If $V_{MTJ} >$  Energy band gap ($ \Delta \left(0 \right)$) and tunneling of MTJ leads to precessional movement of $\vec{m_z}$. The precessional state is where the $\vec{m_z}$ keeps on oscillating at random state (1) showing the impact of voltage biasing and band gap. Adjusting $V_{MTJ}$ helps to reduce the $I_c$ of MTJ's, which will be useful for tuning the device for offset developed after manufacturing as well as to reduce the quantization noise by adjusting the $I_c$. The 3-bit quantizer is shown in  Fig. \ref{fif4}, consisting of 7 SOT-MTJ for producing 7-bit thermometer code. 

%
%

\section{Achirtecture of Spin Flash ADC}\label{section3}
The conventional Spin Flash ADC\cite{spinADC7} has two 3-bit SOT-MTJ, which act as a conversion set as well dummy set. The conversion set is the set where the input is given and MTJs change state according to input value. The dummy set are the one that acts as a reference block for the comparator to compare, so that the comparator can compare between high resistance (AP state) and low resistance (P state). The conventional approach consists of three steps in the conversion of analog to digital. The first step is to quantize the input sampled signal, conversion set SOT-MTJ changes state if the input signal is greater than its own critical current. The second step is to sense the change in resistance of each MTJ by comparing it with a dummy set of SOT-MTJ, so the charge steering StrongARM latch comparator amplifies and converts the change to thermometer code. 
\begin{figure*}[!t]
    \centering
        \includegraphics[scale=0.38]{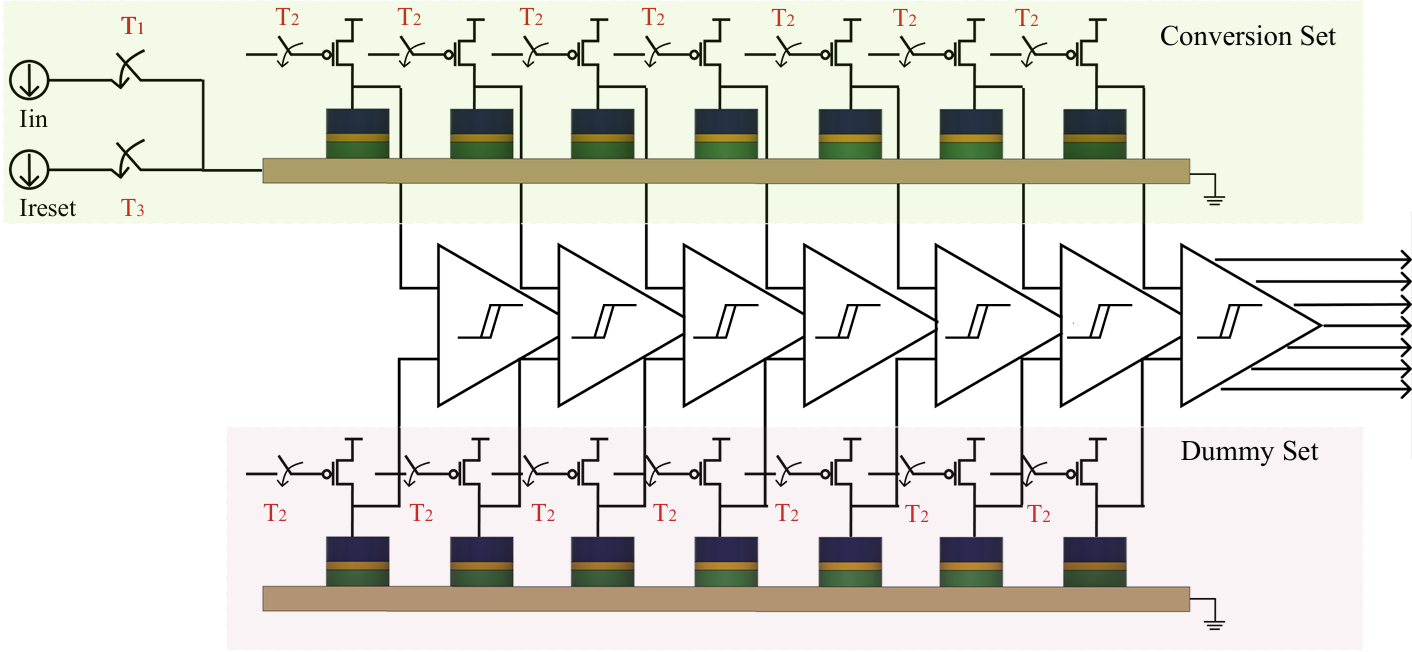}`
	    \caption{Schematic diagram of conventional 3-bit Spin Flash ADC.}
        \label{reference_3bit}
\end{figure*}
\begin{figure}[!t]
    \centering
    \includegraphics[width=\columnwidth]{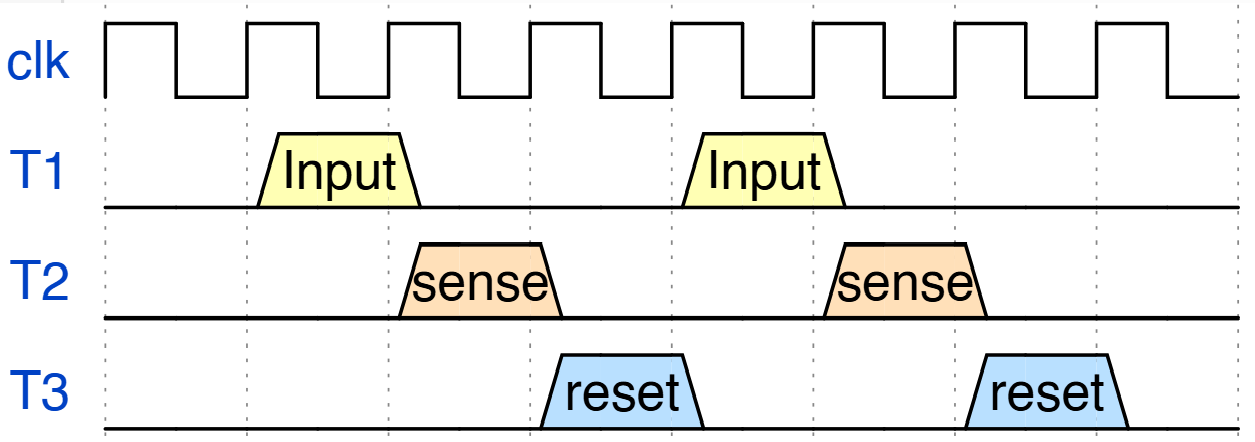}
    \caption{Timing Diagram for conventional 3-Bit Spin Flash ADC}
    \label{fig:TimingDiagram_1}
\end{figure}
In the third step resetting phase, the SOT-MTJ which have switched to AP state are put back into P state as seen in the Fig.\ref{reference_3bit}. The 3 steps all together take 5ns, where conversion and resetting take 4ns and sensing take 1ns thus conversion rate is 200MS/s. The conversion and resetting in the conventional spin flash ADC take 6ns, here by optimizing the timing of SOT, STT, and VCMA effects switching is fastened up. 

\begin{table}[!t]
\caption{Parameters taken for SOT-MTJ model}
\label{tab1}
\centering
\footnotesize
\setlength{\tabcolsep}{3pt}
\begin{tabular}{ll}
\toprule
Parameter & Value \\
\midrule
$\gamma$ (Gyromagnetic ratio) & $2.2127 \times 10^5\,\mathrm{m/(A\cdot s)}$ \\
$\mu_0$ (Vacuum permeability) & $1.2566 \times 10^{-6}\,\mathrm{H/m}$ \\
$k_B$ (Boltzmann constant) & $1.38 \times 10^{-23}\,\mathrm{J/K}$ \\
$q$ (Elementary charge) & $1.6 \times 10^{-19}\,\mathrm{C}$ \\
$\hbar$ (Reduced Planck constant) & $1.054 \times 10^{-34}\,\mathrm{J\cdot s}$ \\
$\alpha$ (Gilbert damping factor) & $0.05$ \\
$M_S(0)$ (Saturation magnetization) & $6.25 \times 10^5\,\mathrm{A/m}$ \\
$K_i(0)$ (Interfacial PMA at $0\,\mathrm{V}$) & $3.2 \times 10^{-4}\,\mathrm{J/m^2}$ \\
$t_{\mathrm{FL}}$ (Free layer thickness) & $1.1\,\mathrm{nm}$ \cite{Voltagegatedmodel} \\
$t_{\mathrm{ox}}$ (MgO thickness) & $1.4\,\mathrm{nm}$ \cite{Voltagegatedmodel} \\
$\omega_{\mathrm{HM}}$ (HM width) & $50\,\mathrm{nm}$ \cite{Voltagegatedmodel} \\
$t_{\mathrm{HM}}$ (HM thickness) & $3\,\mathrm{nm}$ \cite{Voltagegatedmodel} \\
$\mathbf{H}_{\text{In-plane}}$ (In-plane field) & $-40$ Oe \cite{MHM} \\
$\xi$ (VCMA coefficient) & $60\,\mathrm{fJ/(V\cdot m)}$ \cite{VCMA} \\
$D$ (Diameter of MTJ) & $50\,\mathrm{nm}$ \cite{Voltagegatedmodel} \\
$A$ (Surface area of MTJ) & $\pi \times 50^2 / 4$ \\
$T_0$ (Room temperature) & $300\,\mathrm{K}$ \\
$T_C$ (Curie temperature) & $750\,\mathrm{K}$ \\
\botrule
\end{tabular}
\end{table}
\begin{figure}[!b]
	\centering
\includegraphics[scale=0.23]{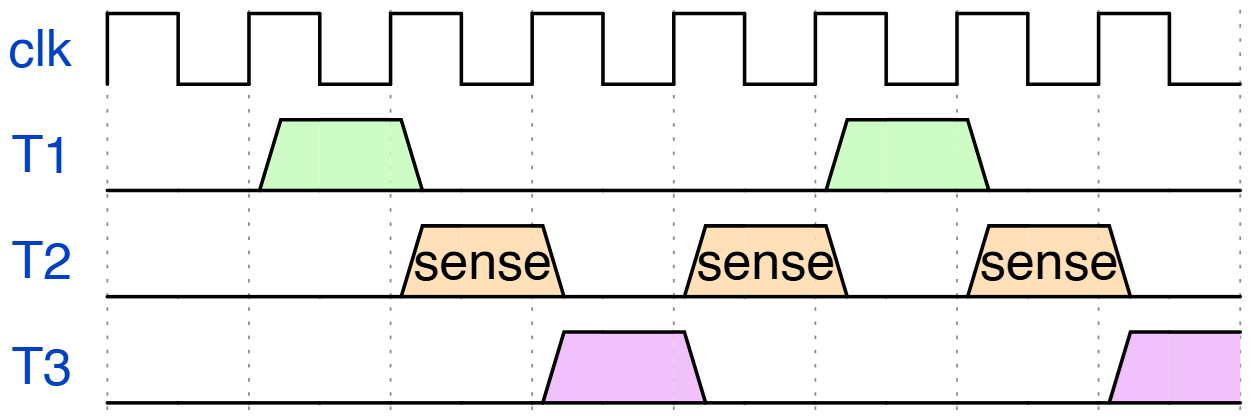}
	\caption{Timing diagram of improved architecture 3-bit Spin Flash ADC. T1 and T3 controls the input and reset currents as seen the proposed block diagram. } \label{Timing_diagram_noval}
\end{figure}
\par
\begin{figure*}[!t]
		\centering
\includegraphics[scale=0.3]{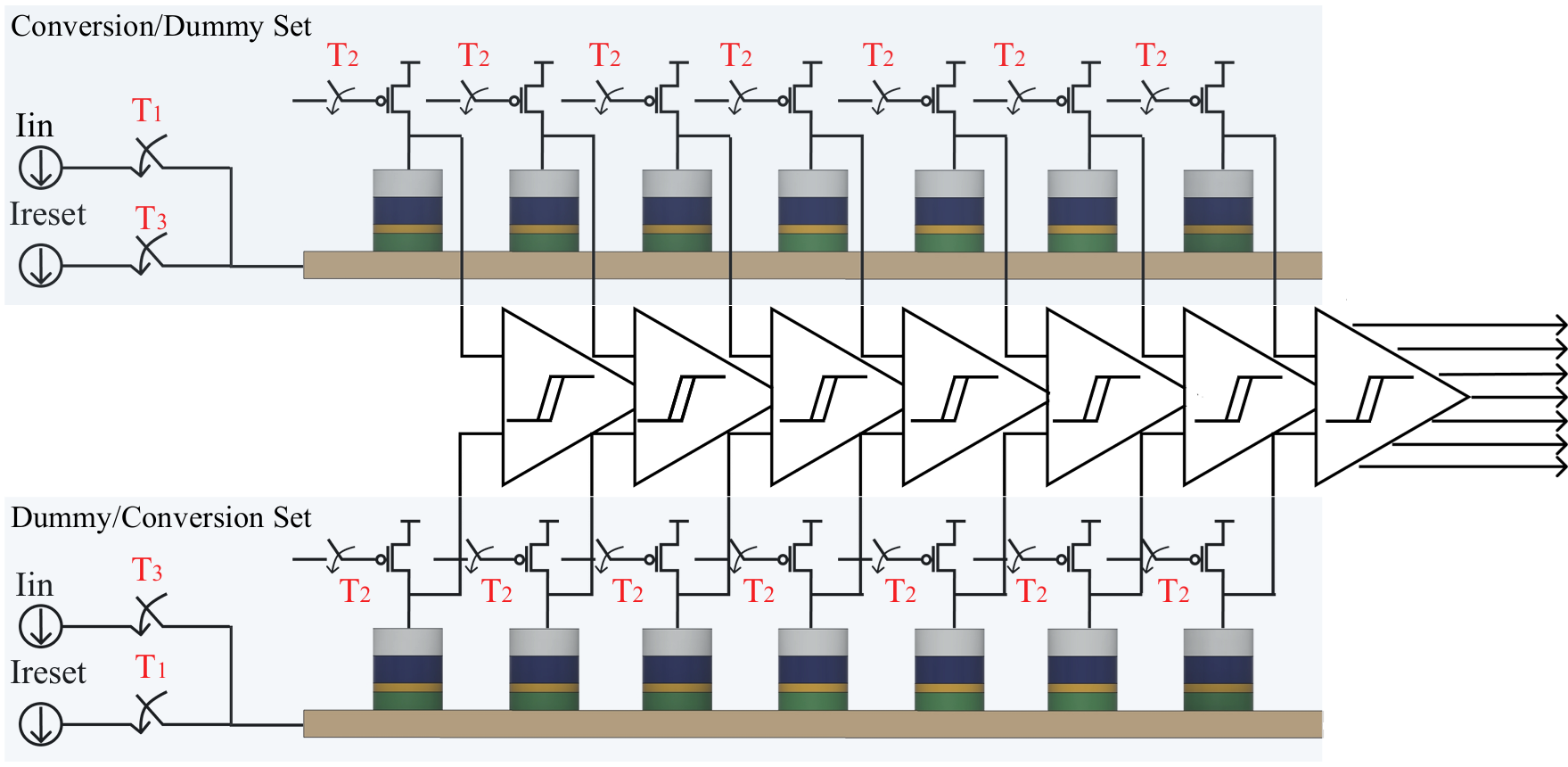}
	\caption{Block diagram of improved architecture 3-bit Spin Flash ADC. The Dummy 3-bit SOT is used for conversion, so the analog to digital conversion step is reduced to two steps.} \label{Noval_3bit}
\end{figure*}

 In the proposed circuit the dummy set and conversion set are interchanged accordingly. Fig.\ref{Noval_3bit} shows that dummy and conversion set of 3-bit SOT-MTJ purposes are interchanged by adding extra switches. Input $I_{SOT}$ and Reset signals are controlled through clock-gated MOSFETs. The timing diagram in Fig.\ref{Timing_diagram_noval} controls the switches for novel architecture. The T1 and T3 signal controls both the Input signal in conversion set and Reset signal in Dummy set and T2 controls the comparator sensing. After the first conversion, simultaneously resetting and conversion occurs.  During the sensing phase voltage difference is captured by the StrongARM latch comparator, the P state has less potential and the AP state has higher potential when compared to the P state. The potential difference between the P and AP states must exceed the noise level from the StrongARM latch comparator. A higher TMR ratio helps suppress the noise and improves reliable detection.
\section{Simulation Results} \label{section4}
\begin{figure}[!b]
\centering
	\includegraphics[scale=0.6]{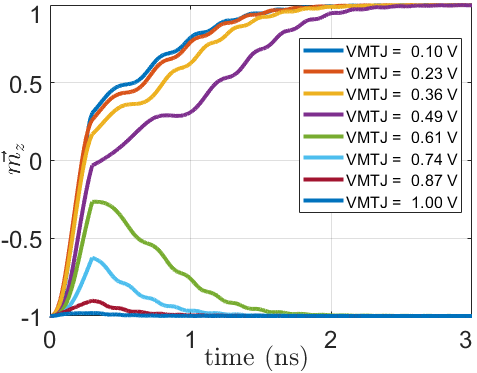}
	\caption {Voltage biasing $V_{MTJ}$ of MTJ reduces $\Delta(Eg)$, which impact switching characteristics and can be used to control the critical current ($I_c$).}\label{fig7}
\end{figure}

The voltage biasing of MTJ plays a vital role in switching from P to AP. Fig. \ref{fig7} shows biasing effects on the MTJ, where the energy barrier gap is reduced. Maximum Voltage biasing $V_{MTJ}$ of 0.4 V is allowed for deterministic switching, for greater voltage MTJ goes to a precessional state and results in stochastic switching. Under negative biasing the energy barrier gap is increased and critical current is increased needs more time as well as SOT current for switching. With intelligent use of voltage biasing the critical current of each MTJ can be changed to capture the input signal and reduce the quantization noise.
\begin{figure}
\centering
	\includegraphics[scale=0.6]{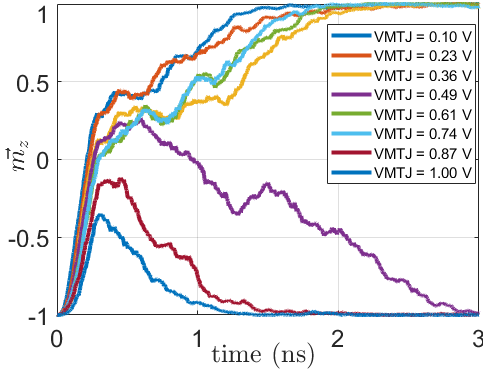}
	\caption {Voltage biasing $V_{MTJ}$ of MTJ including thermal noise produces more stochastic switching.}\label{fig8}
\end{figure}
\begin{figure}[!t]
	\includegraphics[width=\columnwidth]{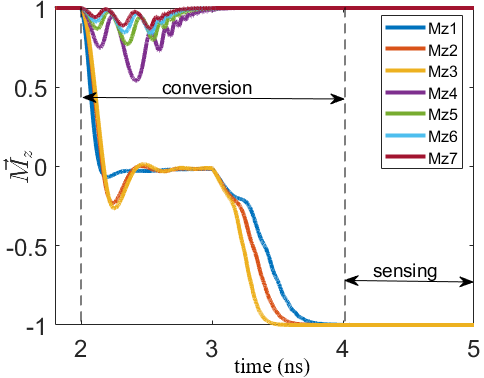}
	\caption{Analpg to Digital conversion of improved architecture 3-bit Spin Flash ADC. }\label{fig11}
\end{figure}\par Thermal noise is an important factor in considering deterministic switching. As voltage biasing varies from -1 to 1 V the magnitude of thermal noise generated is 0 to 2.5 K Oe and the direction is random, which is amble amount to produce deterministic switching to stochastic switching shown in Fig.\ref{fig8}. After SOT current is applied and $\vec{m_z}$ is in an in-plane direction, at this position thermal noise can be dominant and can change the course of switching. Under the biasing MTJ below 0.4, the effect of thermal noise has produced only 8 errors in 100 switchings taken for an MTJ with 50 nm width. \par The proposed circuit is improved by alternatively changing the dummy set and conversion set and removing resetting time thus reducing conversion time, with additions of switches.  Fig. \ref{fig11} shows the proposed circuit $\vec{m_z}$ dynamics, where a 2ns delay is removed which is removing the resetting phase. Charge steering StrongARM latch senses the change in the MTJ state and pushes the output logical one or zero. StrongARM latch is compared with the P state of the dummy set of MTJ and conversion set MTJ. Charge steering in StrongARM lath reduces the power consumption and has 1 GHz as operating speed.
\begin{table}[ht]
\caption{Work Comparison with other Spin Flash ADC}\label{tab2}%
\begin{tabular}{@{}llll@{}}
\toprule
 Parameters & This Work  & \cite{spinADC7} & \cite{ADC1}\\
\midrule
Technology    & Spin-CMOS  & Spin-CMOS & CMOS  \\
&$(45 \mathrm{~nm})$  & $(180 \mathrm{~nm})$  &$(90 \mathrm{~nm})$ \\
Architecture    & Flash   & Flash  & Flash \\
Field Free & Yes & No & - \\
Resolution & 3-Bit & 3-Bit & 3-Bit \\
Conversion   \\ Rate $(\mathrm{S/s})$& $304.1 \mathrm{M}$ & $102 \mathrm{M}$ & $2 \mathrm{G}$  \\
Power (mW) & 0.476 & 0.416 & 3.9 \\
Input Range & 20$\sim$160 ${\mu}A$ & 58$\sim$464 ${\mu}A$& 400(mV) \\
DNL(LSB)& -0.383$\sim$0.245  & -0.275$\sim$258 & -0.24$\sim$0.15 \\
INL(LSB) & -0.149$\sim$0.233 & -0.25$\sim$0.04& -0.24$\sim$0.15 \\
\botrule
\end{tabular}
\end{table}\par Table \ref{tab2} shows the comparisons of the results with past work on Spin Flash ADC and CMOS Flash ADC, where this work shows a three times increase in speed and a slight reduction of power. The CMOS Flash ADC \cite{ADC1} have a current mirror circuit for copying the reference current to each current comparator as well as input current is also copied to each current comparator. The current mirror circuit area increases as the width of MTJ increases, the real advantage of the use of MTJ is eliminating current mirrors and also saving leakage current of the current mirror. Spin-CMOS Flash ADC \cite{spinADC7} have deterministic switching and produced a conversion speed of 102 MS/s after optimizing the action of SOT and STT, which can improved by utilizing dummy SOT-MTJ and speed is improved by 3 times. 


\section{Conclusion}
In this work, a 3-bit spin flash ADC is proposed with improved switching and conversion speed. Analog input is quantized into digital output by SOT-MTJ, with a varying width of HM to produce different critical currents. The impact of voltage biasing on MTJ changes the critical current, which can be used to increase or reduce the critical current. Thermal noise have the ability to change the course of the switching and the design of the stability factor is crucial of deterministic switching. The proposed circuit optimizes the circuit and improves conversion speed. In conventional CMOS ADC, where Vref or Iref can be tunned for better quantization according to application, in work shows variations of critical current with the impact of voltage biasing the MTJ. The Spin Flash ADC works with a conversion time of 3.28ns and power consumption of 476$\mu$W, respectively.
\subsection*{\textbf{Author Contributions}}
N.K. conceived the conceptualization and formalization of research and coordinated the project.  A.F. performed numerical modelling of field-free magnetic tunnel junctions through the supervision of N.K. A.F wrote the initial draft. A.F, N.K and P.p analyzed the data and revised the manuscript.

\section*{Declarations}

\textbf{Conflict of Interest} The authors have no relevant financial or non-financial interests to disclose.

\balance
\bibliographystyle{sn-mathphys.bst}
\bibliography{sn-bibliography}

\end{document}